\documentclass[twocolumn,preprintnumbers,amsmath,amssymb]{revtex4}
\usepackage{graphicx}
\usepackage{amsmath}
\usepackage{amssymb}
\usepackage{natbib}
\usepackage{subfigure}
\usepackage{color}
\usepackage{lipsum}
\usepackage[T1]{fontenc}
\usepackage[final]{pdfpages}

\begin{document}
 
\title{Crucial role of non-specific interactions in amyloid nucleation}

\author{An\dj ela~\v{S}ari\'c$^1$, Yassmine~C.~Chebaro$^{1}$, Tuomas~P.~J.~Knowles$^{1}$ and 
Daan~Frenkel$^{1}$}
\email{df246@cam.ac.uk}
\affiliation{$1$Department of Chemistry, University of Cambridge\\ Lensfield Road, Cambridge CB2 1EW\\ }

\begin{abstract}
Protein oligomers have been implicated as toxic agents in a wide 
range of amyloid-related diseases. Yet it has remained unsolved whether the oligomers are a necessary step in the formation of amyloid fibrils, or just a dangerous 
by-product. Analogously, it has not been resolved if the amyloid nucleation 
process is a classical one-step nucleation process, or a two-step process 
involving pre-nucleation clusters.  We use coarse-grained computer 
simulations to study the effect of non-specific attractions between peptides on 
the primary nucleation process underlying amyloid fibrillization. We 
find that for peptides that do not attract, the classical one-step nucleation 
mechanism is possible, but only at non-physiologically high peptide concentrations.  At low 
peptide concentrations, which mimic the  physiologically relevant regime, 
attractive inter-peptide interactions are essential for fibril formation.  
Nucleation then inevitably takes place through a two-step mechanism involving prefibrillar 
oligomers. We show that oligomers not only help peptides meet each 
other, but create an environment that facilitates the conversion of monomers 
into the $\beta$-sheet rich form characteristic of fibrils.  Nucleation typically does not proceed via the most prevalent oligomers, but via an oligomer size that is only observed in rare fluctuations, which is why such 
aggregates might be hard to capture experimentally. Finally, we find that the 
nucleation of amyloid fibrils cannot be described by classical nucleation 
theory: in the two-step mechanism the critical nucleus size increases both with 
an increase in concentration and in the inter-peptide interactions, in direct 
contrast with predictions from classical nucleation theory.
\end{abstract}


\maketitle

\section*{Significance Statement}
The assembly of normally soluble proteins into large fibrils, known as amyloid aggregation, is associated with a range of pathologies. Prefibrillar protein oligomers, not the grown fibers, are believed to be the main toxic agents. It is unresolved if these oligomers are necessary for fibril assembly, or just a dangerous by-product. We show by using computer simulations that at physiological concentrations, amyloid formation must proceed via a two-step process including prefibrillar oligomers. We find that there is an optimal oligomeric size for amyloid nucleation, and that classical nucleation theory cannot be applied to this process. Formation of oligomers and hence fibrils, is controlled by the strength of non-specific attractions, whose weakening may be crucial in preventing amyloid aggregation.
\bibliographystyle{pnas2011}

\section*{Introduction}
During the process of amyloid formation, normally soluble proteins assemble into 
fibrils that are enriched in $\beta$-sheet content, and have diameters of a few 
nanometres and lengths up to several microns.  This phenomenon has been 
implicated in a variety of pathogenic processes, including Alzheimer's and 
Parkinson's diseases, Type II diabetes and Systemic 
Amyloidoses~\cite{selkoe,dobson,chiti}. The association with human diseases 
has largely motivated a long standing effort to probe the assembly 
process and numerous studies have aimed at elucidating the mechanism of 
amyloid aggregation~\cite{radford}. The basic nature of the aggregation reaction 
has emerged as a nucleation and growth process~\cite{jarrett,dobsonrev}, where the aggregates are created 
through a not well understood primary nucleation event, and can grow by recruiting further peptides or 
proteins to their ends~\cite{oosawa1962theory, Eaton}. In this paper 
we focus on the nature of this primary step in amyloid nucleation and the fundamental initial 
events that underlie amyloid formation. 

Amyloidogenic peptides and proteins when in their non-pathological cellular form can range in the structures from mainly $\alpha$-helical, $\beta$-sheet and even random coil, while the amyloid forms of proteins 
possess a generic cross-$\beta$ structure~\cite{fitzpatrick,wasmer,serpell,tycko,fandrich,eisenberg}.  The formation of amyloid is hence accompanied by 
marked changes in the conformations of the peptides and proteins that undergo 
this process. A pertinent question is whether this conformational change 
takes place simultaneously with the nucleation process or whether nucleation 
takes place first and is then followed by conformational change. These two 
possible scenarios of nucleation have been extensively discussed in the experimental and 
theoretical literature~\cite{jarrett,lomakin,Eaton,auer,lindquist,kelly,linse}. We will refer in this work to the two scenarios simply 
as the one-step nucleation (1SN), in which the $\beta$-sheet enriched nucleus 
forms directly from the solution, and the 
two-step nucleation (2SN), where soluble monomers first assemble into disordered 
oligomers, which subsequently convert into a $\beta$-sheet 
nucleus. Disordered oligomers, ranging in size 
between dimers and micron-sized particles, have been observed in some 
experiments~\cite{garzon,shankar,liang,bernstein,bernstein1,bitan,bleiholder,sabate,yong}.  
These findings highlight a central question regarding the role of disordered 
oligomers in fibril formation: are such clusters a necessary step in the 
process of fibril formation, or are they just a by-product?


From a biological and biomedical perspective, it is important to understand the 
conditions under which oligomeric clusters form because such species exhibit 
high cytotoxicity~\cite{kenjiro, hardy2002amyloid, selkoe, bucciantini}.
Indeed, there is strong evidence that the disordered oligomers, rather than 
fully-grown fibrils, are the main pathogenic species in protein aggregation 
diseases~\cite{haass, bucciantini,walsh}. As such, defining the role of the 
prefibrillar oligomers during amyloid formation will be crucial to develop 
intervention strategies that target these species~\cite{Hefti2013b, lansbury2006century, selkoe, hardy2002amyloid}.

Mutations in the polypeptide sequence and extrinsic changes in the experimental conditions are known to alter the concentrations of aggregated species, their 
size and cytotoxicity~\cite{bitan,carullaa,crescenzi,melchor,rebeck}. For 
instance, mutations that increase hydrophobicity of the Alzheimer's beta peptide 
A$\beta$1-42 have a pronounced effect on its 
aggregation behaviour and the size distribution of the resulting 
oligomers~\cite{bitan,urbanc2,bernstein,bernstein1,bleiholder}, promoting 
toxicity and expediting the fibrillization process. In the same spirit,
two extra hydrophobic residues in A$\beta$1-42 are believed to contribute to the more pronounced oligomerisation and faster fibrillization compared to its alloform A$\beta$1-40~\cite{bitan,bernstein1,urbanc2}.  Temperature, pH, and concentration of certain metals also affect oligomerization and pathways of fibrillization~\cite{jiang,bhowmik,gorman,srinivasan}.\par
The common feature of above experiments is that they modify the internal free energy difference between the soluble and the $\beta$-sheet forming state, also called the $\beta$-sheet propensity, which has been extensively studied in the literature~\cite{pellarin,sheajcp,thirumalaiprl,shearef}. However, they also modify interactions between peptides that aggregate, a crucial contribution that has not yet been systematically addressed.

In this paper we study the effect of non-specific interactions between peptides 
on the amyloid nucleation process. Such non-specific interactions do not depend 
on the atomistic details of the amino acids involved, allowing us to 
address question about amyloid aggregation and nucleation, using a 
coarse-grained model. In particular, generic hydrophobic stretches in the 
sequence of A$\beta$, have been shown to be sufficient to promote 
aggregation~\cite{hecht1,hecht2}. Mutations of nonpolar residues to other 
nonpolar residues had a little or no effect on aggregation, while mutations that 
reduce charge and/or increase hydrophobicity enhanced it~\cite{hecht2,klimov}. Furthermore, 
atomic force microscopy 
measurements have shown that the strength of overall  interactions between 
amyloidogenic proteins correlates with their tendency to 
aggregate~\cite{lyubchenko1,lyubchenko2}.

We have performed extensive computer simulations that allow us to observe both 
the 1SN and the 2SN mechanisms. These simulations reveal that 1SN and 2SN can be 
viewed as two limits of the same process, something that several previous studies have 
suspected~\cite{kelly,auer}. Importantly, we observe that only 2SN is possible 
at low peptide concentrations, comparable to the levels that are found in vivo.
Another key observation is that fibril nucleation typically does not proceed via 
the most prevalent oligomeric species, but rather via an oligomer with a size 
that is only observed as a result of rare fluctuations.  As a consequence, such 
oligomers will be hard to capture experimentally, although their presence 
is required for nucleation to take place.  Our simulations show that the free 
energy barrier for fibril nucleation via the two-step mechanism decreases with 
increasing strength of the inter-peptide interactions. Furthermore, the critical 
nucleus size in the two-step mechanism is found to grow  with the increase in 
the peptide concentrations as well as with stronger interpeptide interactions, 
which is in direct contrast with the classical nucleation. These results imply that weakening the non-specific interactions between peptide monomers in solution, and thereby simultaneously increasing both the free energy barrier for oligomer formation, and the free energy barrier for peptide conversion at a given oligomer size, may be a crucial step in preventing amyloid aggregation.

\textbf{Coarse-Grained Model of an Amyloidogenic Peptide Supported by Atomistic Simulations.} In view of the experimental evidence that the formation of peptide oligomers is 
governed by generic features of the interactions between 
monomers~\cite{hecht1,hecht2,dobson3}, we can employ a simple, coarse-grained, model of an amyloidogenic peptide. The great advantage of such a model is that it is sufficiently cheap to  allow us
 to vary inter-peptide interactions and explore a wide range of peptide concentrations. The model should be able to capture both the formation of amorphous oligomers and the nucleation of fibrils.  Earlier studies~\cite{vacha} have shown that a minimal model that captures this phenomenology accounts for the fact that the peptides can be in two states:
 a soluble state that can form only disordered oligomers, and a higher free-energy state  that can form $\beta$-sheet-like fibrils 

The soluble state of the amyloidogenic peptide is modelled as a hard spherocylinder with an attractive patch at the tip (Fig. 1A). The patch is the source of non-specific attractions, whose strength is represented by the parameter $\epsilon_{ss}$. 
Such particles are able to make micellar-like oligomers, as represented in Fig. 1C, \textit{Upper}, but not extended aggregates. The $\beta$-sheet forming configuration is described as a hard spherocylinder with an attractive side-patch (Fig. 1A). Peptides in the $\beta$-state interact with an interaction strength $\epsilon_{\beta\beta}$ if their patches point towards each other. In that case they tend to pack parallel to one another, which leads to the fibril-like structure (Fig. 1C, \textit{Lower}). 
The interaction strength between the soluble and the fibril-forming state is given by $\epsilon_{s\beta}$ (Fig. 1A). In our Monte Carlo simulations  a soluble protein can convert into the $\beta$-prone state every simulation step with some small probability that mimics slow conversion of the soluble peptide into the $\beta$-prone configuration. The conversion, apart from being kinetically slow, is also thermodynamically unfavourable, which reflects the loss of the conformational entropy of the $\beta$-prone form compared to the soluble form~\cite{eisenberg}. Hence we penalize every conversion from the soluble to the $\beta$-form with a change in the excess chemical potential of $\Delta\mu=20\textrm{k}T$, where $\textrm{k}$ denotes Boltzmann's constant. This value is chosen to reflect the fact that amyloidogenic proteins are typically not found in the $\beta$-sheet conformation in solution ~\cite{dobson3,vendruscolo}. With respect to the current body of work on the importance of the $\beta$-sheet propensity, this places our model in the range of proteins with small- to mid-$\beta$-propensity, such as A$\beta$. Further details of the coarse grained model can be found in the Supporting Information I.

To obtain an estimate of the interaction parameters  $\epsilon_{\beta\beta}$, $\epsilon_{s\beta}$ and  $\epsilon_{ss}$, we carried fully atomistic simulations in the system of the A$\beta$1-42 peptide in explicit water, and used umbrella-sampling to determine the three possible pair-interactions. Details of the simulation procedure are explained in SI Text, section II and Fig. S1. As expected, the interaction between two peptides in the hairpin conformation packed in the $\beta$-sheet, reflecting the $\epsilon_{\beta\beta}$ interaction, is the strongest ($-15\textrm{k}T$), as shown in Fig. 1B. To obtain an estimate of the interaction between the soluble and the $\beta$-prone peptide, we constrained one of the peptides to be in the U-turn shape of a $\beta$-hairpin configuration and sampled its interaction with the random-coil peptide. Although the atomistic properties of the isolated $\beta$-prone peptide are still unknown, our purpose here is to use a conformation that can easily act as a platform for fibril assembly. We find that the interaction energy of this $s-\beta$ complex at the shortest sampled distance is $-8.7~\textrm{k}T$. The interaction between the peptides in random coil conformation is the smallest of the three ($-5.4~\textrm{k}T$), but still clearly attractive. In our model, the strength of this attractive interaction is represented by $\epsilon_{ss}$. 
It is important to note that the shortest sampled interpeptide separation in each of the three cases is limited
by the sum of the peptides' radii of gyration.
Of course, it is an oversimplification to try to capture the interaction energy between two complex, fluctuating peptides by a single number. However, we argue 
that the relative interaction strengths that we compute in our atomistic simulations are indicative of the relative strengths of the interactions in  the first steps of assembly. 
We use the atomistic simulations mainly to justify why, in our coarse-grained model, we choose the following ordering of interaction strengths: $\epsilon_{\beta \beta} > \epsilon_{s\beta}  > \epsilon_{ss}$.
An additional set of umbrella simulations was performed on the A2V mutation of A$\beta$42, which is reported to be responsible for early-onset Alzheimer's disease~\cite{Fede2009}, with the purposes of evaluating the relative strengths of interactions for an additional system and further supporting the choice of energy scale in the coarse-grained model. The A$\beta$42-A2V simulations
show the same trend for the interaction strengths as the wild-type (WT), with 
$\epsilon_{\beta \beta} > \epsilon_{s\beta}  > \epsilon_{ss}$ ($14.6$, $5.9$ and $4.3\textrm{k}T$ respectively (Fig. S2 (D)). (We do not discuss implications of this mutation on the mechanism underlying the A2V substitution in A$\beta$ fibrillogenesis. To accurately compare it to the wild-type, we would also need to asses the free energy barrier for the $\beta$-conversion associated with both variants, which is not within
the scope of this work.)

\textbf{1SN \textit{Vs.} 2SN.} Our Monte Carlo simulations allow us to study amyloid nucleation pathways over a 
range of peptide concentrations and interpeptide interaction strengths.
A tentative phase diagram is shown in Fig. 2A. When interactions between peptides 
in the solution are sufficiently strong ($\epsilon_{ss} \gtrsim 5 \textrm{k}T$), 
disordered, micellar-like oligomers are formed before the observation of 
elongated fibrils.  These oligomers exist in the solution for a long time before transforming into a $\beta$-sheet nucleus through a series of single-peptide conversions, during which each converted peptide is in contact 
with other random coils within the aggregate.  This two-step mechanism (2SN) is 
often referred to in the literature as the nucleated-conformational conversion 
(NCC), and was observed both for $A\beta$ peptides~\cite{kelly} and prion 
proteins~\cite{lindquist} under certain experimental conditions. The $\beta$-sheet protofibril can 
further grow by either addition of monomers from the solution or merging 
with another oligomer. The preferred mode of growth will depend on the exact 
conditions: at lower concentrations and lower $\epsilon_{ss}$-values, the system 
is well below the critical micelle concentration, and there will be
many more monomers than oligomers in solution~\cite{israelachvili}. In that case, the growth will 
mainly be through monomer addition~\cite{lindquist}. Above the critical micelle concentration, 
most of peptides will be incorporated within oligomers, and few monomers will 
remain in solution. In that case, the addition of disordered oligomers at the 
end of the growing fibre becomes the more probable fibril growth mechanism~\cite{kelly}.\par 
Our atomistic simulations show that the 
attractions between soluble peptides originate mainly 
from polar interactions, with some participation of hydrogen-bonding 
(Fig. S2 B and C). Recent molecular simulations support this observation and indicate that the 
polar N-terminal region can act as a catalytic region that initiates and accelerates assembly of A$\beta$ 
peptides~\cite{Chong2012,Viet2014}.\par
If the interactions between peptides are weak ($\epsilon_{ss} \lesssim 5 \textrm{k}T$), 
we observe that fibril nucleation can happen only when the peptide concentration 
is high - above 2 mM in our case. At high concentrations, random fluctuations 
can push monomers close to each other, and a nucleus can be born from the 
solution in a single step without any noticeable preceding aggregate (1SN). 
This event is followed by the rapid growth by monomer addition and sometimes 
referred to as the nucleated-polymerisation (NP) ~\cite{Eaton,jarrett}. It is 
important to point out that the lifetime of the disordered oligomers is the 
key feature that distinguishes this mechanism from the 2SN route discussed 
above. Fig. 2B, \textit{Left} shows the representative snapshots during the 
1SN nucleation process. Closer inspection reveals that, because of random 
fluctuations, several soluble peptides interact with each other and other 
peptides that can convert to the $\beta$-prone state because of the stabilizing 
interaction with the soluble peptides. Interpeptide interactions are 
crucial even in this case, but the event is transient: it does not require the 
presence of long-lived oligomers. (Long-lived in our simulations means that, after oligomers form, they tend to stay stable for most of the simulation time.)  We should emphasize that the 
cross-over from 1SN to 2SN is not abrupt, and in the regime of high peptide 
concentrations and intermediate interactions, one can observe nucleation from 
short-living oligomers, which exhibits features of both limits.\par
It is worth adding that at much higher peptide concentrations and interpeptide interactions than the ones reported here, kinetically arrested amorphous aggregates can be observed~\cite{dima,boden}. We did not study this region, since we focus on more physiologically relevant concentrations and interaction values.\par
\textbf{Nucleation at Low Peptide Concentrations.} 
The most relevant regime for amyloidogenic peptides under physiological 
conditions is the regime of low peptide concentrations (micro- and nano-molar). 
Our simulations clearly show that in this regime 1SN is extremely unlikely - so 
much so that we did not observe a single 1SN event in our longest ($10^9$ Monte Carlo 
steps)
simulation runs. The explanation is simple: in this regime, the probability of a random encounter of enough monomers to initiate fibril nucleation is  completely negligible. At low peptide concentrations nucleation can only be achieved if there are appreciable attractive interactions between soluble monomers.  
Such nonspecific attractions favor the formation of long-lived peptide clusters in which subsequent fibril nucleation can take place.
 For instance, at a concentration of $c \sim 70 \mu M$, we observe fibrils above the threshold of $\epsilon_{ss} \gtrsim 8\textrm{k}T$. In this case, nucleation originates inside 
oligomers that form only rarely but are relatively long-lived. Fig. 2B, \textit{Right} shows an example of the 2SN mechanism that we observe.
Direct experimental observation of the first stages of 2SN may be challenging as the concentration of the oligomers is very small.
However, our simulations suggest that it should be straightforward to obtain indirect evidence of the 2SN mechanism: proteins that do not attract in solution, and cannot form prefibrillar clusters, are very unlikely to form fibrils at low concentrations.\par

\textbf{Role of Oligomers in the Nucleation Process: Free Energy Analysis.} Non-specific attractions between soluble peptides favour the formation of relatively long-lived oligomers.  The existence of oligomers lowers the nucleation barrier for the subsequent fibril formation because, as our atomistic simulations show,  the  $s\beta$-interaction is more favourable than the $ss$-interaction, and the $\beta\beta$-interaction is more favourable still. Hence, the energetic price to pay for the conversion from the $s$-state to the $\beta$-state inside an oligomer is compensated by the gain in interaction energy between the peptides. Increased number of hydrophobic contacts between the $\beta$-form and the soluble peptide is the reason why the $s\beta$-interaction is favourable (Fig. S2A). This enhanced hydrophobic  interaction  stabilizes the amyloidogenic $\beta$-form when in contact with other random coils within a disordered aggregate. To quantify this effect, we performed Monte Carlo umbrella sampling simulations to compute the free energy barrier for fibril formation inside an oligomer in our coarse-grained system. Details of the simulation method are given in SI Text, section II and Fig. S1. 

We find that the first conversion step (i.e. the one involving the first 
soluble monomer converting to a $\beta$-form within the oligomer) determines 
the height of the free energy barrier. The finding that the first $s-\beta$ 
conversion has the highest free energy cost is not too surprising: in this 
process, the conformational entropy of one random coil is lost, and there is 
only a small gain caused by interactions of the $\beta$-form with the remaining 
random coils. As soon as the second monomer converts to the $\beta$-form, a 
$\beta$-sheet interaction is established, which is enthalpically very favourable 
because of the increase in hydrophobic contacts and hydrogen bonds (see Fig.
S2 A and C).
Every subsequent contact between the $\beta$-sheets lowers the free energy of the oligomer even more, thus the rate limiting step is the formation of the first $\beta$-form peptide. Hence, in our simulations the critical nucleus for fibrilisation is an oligomer (of yet undetermined size) with exactly one converted peptide. The smallest possible fibril is then made of two $\beta$-peptides, while in practice such small fibrils quickly continue growing. \par 
Our umbrella sampling simulations allow us to calculate the free energy change for converting one peptide into the $\beta$-state within an oligomer of the size $N$, when varying the $\epsilon_{ss}$ peptide interactions (Fig. 3A). We find that the barrier for the conversion $\Delta F_c(N)$ systematically decreases with increasing size of the oligomer for a fixed $\epsilon_{ss}$ interaction. Such behaviour is to be expected as the number of possible $s\beta$-interactions increases with the oligomer size. Of course, the effect of the oligomer size eventually saturates when the maximum number of geometrically allowed interactions is reached.

Not surprisingly, the nucleation barrier also systematically decreases with 
stronger interpeptide interactions. This finding can be seen to emerge from 
the assumption that $\epsilon_{s\beta}$ shifts together with $\epsilon_{ss}$: 
strong $\epsilon_{ss}$ is compensated for by strong(er) $\epsilon_{s\beta}$. To 
inspect the origin of this correlation, we measured $\Delta F_c(N)$ for different 
values of s-$\beta$ stabilizations (Fig. S3) and found that the 
nucleation barrier drops down dramatically with stronger $\epsilon_{s\beta}$ 
interaction.  
This effect is likely to contribute to the 
higher rate of nucleation of $A\beta$1-42 compared to $A\beta$1-40~\cite{MeislPNAS}, and 
peptides with hydrophobic mutations compared to their respective WTs. 
However, at very high values of $\epsilon_{ss}$ (over 10 $\textrm{k}T$), the oligomer reorganization might become very slow, and the process can be kinetically arrested. 
Despite the 
fact that the exact number of converted peptides that give rise to nucleation 
can be dependent on the simulation model or the protein under study, and is 
governed by the $\epsilon_{\beta\beta}/\epsilon_{ss}$ ratio, the trends 
described here should be general. \par
\textbf{Two Competing Free Energy Contributions Cause Nonclassical Scaling of the Critical Nucleus Size.} The overall nucleation barrier for fibril nucleation is the sum of two contributions: the first is the free-energy cost of forming an oligomer of size $N$. The second is the free energy cost of nucleating a fibril in such an oligomer. Both free energy barriers depend on $N$. We can compute  $\Delta F_o(N)$, the free energy barrier for the formation of an oligomer of size $N$ from measurements of the oligomer size distribution, $P(N)$, using  $\Delta F_o(N)=-\textrm{k}T\ln  P(N)$.Examples of $P(N)$ measurements for several values of $c$ and $\epsilon_{ss}$ are shown in Fig. S4. For dilute solutions of clusters, $P(N)$ is of the form $P(N)\sim e^{-\Delta f(N)/\textrm{k}T} c^N$, where $\Delta f(N)$ is a concentration-independent free-energy and $c$ is the monomer concentration. Clearly, for low values of $c$, 
$\Delta F_o(N)$ will grow rapidly with cluster size, which has important consequences for the overall behaviour of the free-energy barrier for fibril nucleation: for low monomer concentrations, the lowest overall nucleation barrier will correspond to nucleation of a fibril in a relatively small oligomer. For higher concentrations, fibril nucleation in larger oligomers will dominate. Fig. 3B shows an  example of the various contributions to the fibril nucleation barrier for one specific choice of the monomer concentration and the value of 
$\epsilon_{ss}$. The shift of the nucleation barrier with monomer concentration is shown in Fig. 4A, while the arrows indicate the respective dominant nucleus sizes. Fig. 4B shows the variation of the nucleation barriers with the non-specific monomer-monomer interaction ($\epsilon_{ss}$). Fig. 4B clearly shows that, all other things being equal, stronger non-specific interactions between soluble peptides tend to decrease the fibril nucleation barrier very substantially. The underlying reason is that $\Delta F_o(N)$ decreases strongly with increasing $\epsilon_{ss}$. We stress that the nucleation of fibrils cannot be described by classical nucleation theory: that theory would predict that  the size of the critical nucleus would grow both with a decrease in  concentration and in the inter-peptide interactions. The explanation is that, unlike the classical nucleation case, we are always working in the regime where the condensed phase of the soluble peptides is thermodynamically unstable. Hence large clusters are disfavoured, in particular at low concentrations. Yet, fibril nucleation must proceed via a cluster, and that process is easier in large clusters (if they can form). It is the compromise between these two trends that lead to the observed variation of the critical nucleus size with monomer concentration and $\epsilon_{ss}$. 

\textbf{Discussion and Conclusions.} Our numerical study reveals the crucial effect of non-specific inter-peptide attractions on the pathways of amyloid aggregation. We find that for  peptides that do not attract, the classical one-step nucleation mechanism is possible, but only at very 
high peptide concentrations. At low peptide concentrations, attractive inter-peptide interactions are essential for  fibril formation. Nucleation then inevitably takes place through a two-step mechanism where the formation of disordered oligomers preceded fibril nucleation. Although direct measurements of such non-specific interactions have been challenging to achieve, the determination of second virial coefficients with advanced light scattering techniques~\cite{chi} is a promising avenue to access these values for amyloidogenic peptides and their mutants in solution.\par
We recall that this study focused on peptides with low- to mid-$\beta$-propensity ($\Delta\mu=20\textrm{k}T$). If the $\beta$-state were stable and could exist in solution, oligomers would not be needed to aid nucleation and the 1SN scenario would become more probable. However, the available experimental data suggest that the $\beta$-state
is usually not present in solution, at least not for the A$\beta$ peptide. It has been observed that mutants that further decrease $\beta$-sheet propensity (such as the E22GA$\beta$) exhibit enhanced oligomer formation over the wild-type~\cite{klimov,nilsberth,lashuel}, which is in agreement with our findings. Our coarse-grained model, although supported by atomistic simulations, is of course highly simplified as we consider only two peptide states. In reality, we expect there to be several intermediate conformations on the pathway between the soluble and the $\beta$-form. Qualitatively, the key physical behaviour originates from the fact that the ``mixed'' interaction energy $\epsilon_{s\beta}$ is intermediate between that of the pure soluble and the $\beta$-form. As such, the present results are expected to generalise to other coarse graining schemes which exhibit this property, including ones with more than one intermediate state.\par
In this framework, oligomers play a dual role. Disordered oligomers not 
only help peptides meet each other, but also create an environment that 
facilitates the conversion of monomers into the $\beta$-form.
We stress that whilst the oligomers that form due to attractive non-specific 
interactions are rare (and therefore hard to observe directly), their formation is a 
crucial on-pathway step in the amyloid formation process for peptides whose $\beta$-competent state is not particularly stable. Similar two-step nucleation processes involving amorphous precursors have been 
identified in very different physical phenomena, such as protein crystal 
nucleation~\cite{tenwolde}, nucleation of sickle-cell haemoglobin 
polymers~\cite{vekilov} and biomineralization~\cite{coelfen}. The surprising 
finding of the present work is that at realistic (i.e. low) peptide 
concentrations, fibril formation {\em must} proceed via an intermediate 
amorphous oligomer and that this pathway leads to highly non-trivial 
predictions for the dependence of the critical nucleus size and nucleation 
barrier on monomer concentration and inter-peptide interaction. 
Since oligomeric species formed under such conditions consist predominantly of non-beta structures, it is interesting to speculate that nature might have evolved a portfolio of chaperones to target specifically both the non-beta as well as the conventional beta aggregated forms to curtail pathological aggregation. Recent evidence suggests that such a situation might indeed occur for extracellular chaperones~\cite{narayan1,narayan2}. Whilst the present simulations focused on the  A$\beta$1-42 peptide, we expect aspects of the mechanism proposed here to be applicable to other amyloid proteins, such as the prion protein or $\alpha$-synuclein. The conformational change from the soluble into the $\beta$-prone state is a ubiquitous feature underlying the amyloid aggregation, and the importance of disordered oligomers could be of a more general nature.\par

\section*{Acknowledgments}
We thank Michele Vendruscolo, Iskra Staneva and William M. Jacobs for helpful discussions. A.\v{S}. acknowledges support from the Human Frontier Science Program and Emmanuel College. Y.C.C. and D.F. are supported by the Engineering and Physical Sciences Research Council Programme Grant EP/I001352/1. T.P.J.K. acknowledges the Frances and Augustus Newman Foundation, the European Research Council, and the Biotechnology and Biological Sciences Research Council. D.F. acknowledges European Research Council Advanced Grant 227758.

 
\begin{figure*}[t!]
\centering
\includegraphics[width=0.95\textwidth]{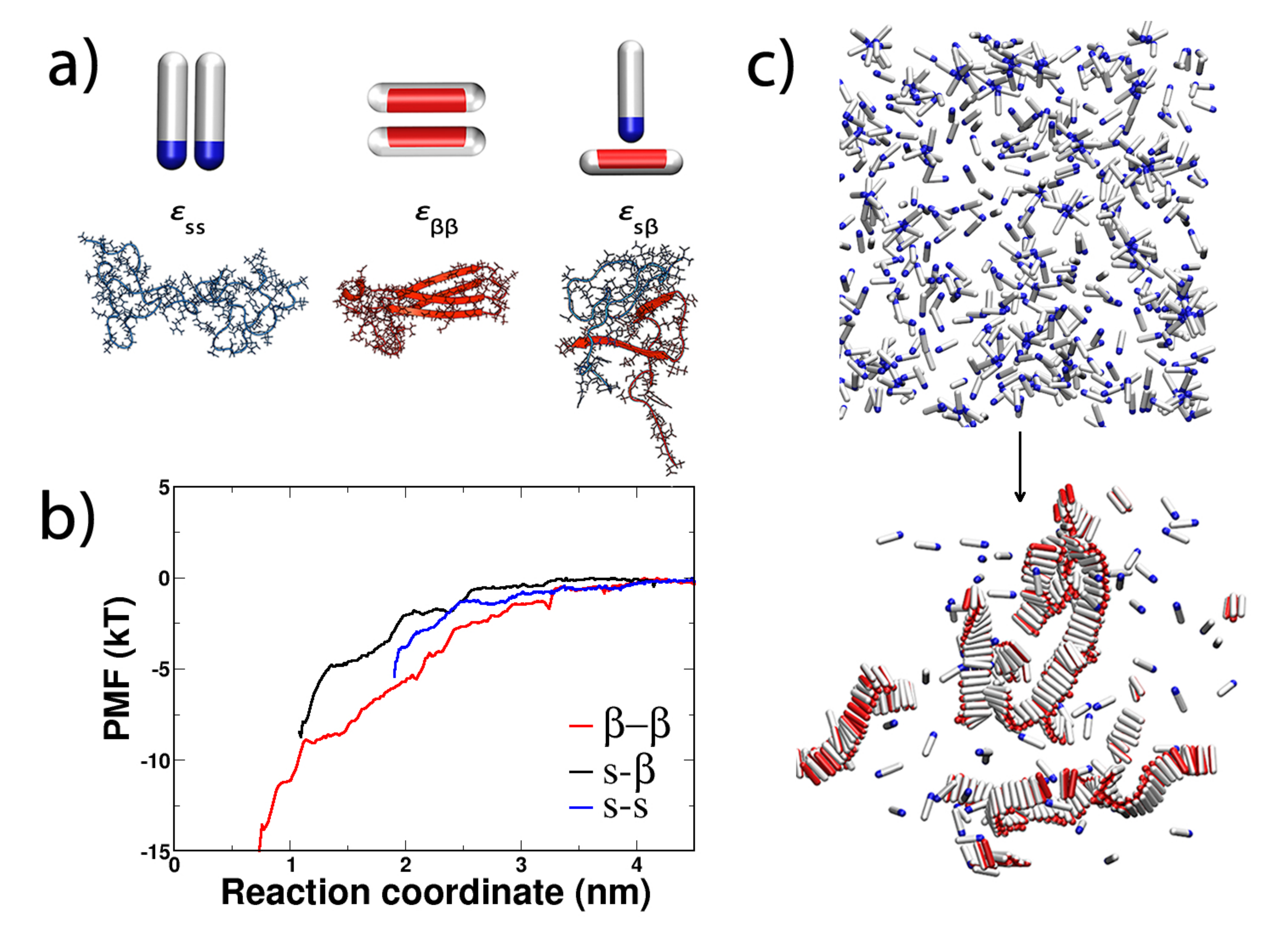}
\caption{Possible interactions in the system. (\textit{A, Left}) Two spherocylinders representing the soluble peptides interact via their attractive tips (blue). The interaction strength is given by $\epsilon_{ss}$. (\textit{A, Center}) Spherocylinders in the $\beta$-form interact via their side-patches (red). The strength of this interaction is $\epsilon_{\beta\beta}$. (\textit{A, Right}) The interaction between the soluble peptide and the $\beta$-competent form, given by $\epsilon_{s\beta}$. Underneath each coarse-grained representation is its atomistic realization for the case of the A$\beta$1-42 peptide. Snapshots are taken at the shortest distance of the Molecular Dynamics umbrella sampling at 10ns. (\textit{B}) Potential of mean force (PMF) for the pair interactions in the A$\beta$1-42 system used to guide coarse-grained simulations. The red curve represents the interaction between two peptides in the $\beta$-form, the blue curve is the interaction between the random-coil peptides, while the black line is the interaction between the random-coil peptide and the peptide kept in the $\beta$-prone form. (\textit{C}) Representative snapshots before the nucleation occured, and at the end of the simulation run. (\textit{Upper}) Disordered oligomers formed by peptides in their soluble form, red circle magnifies  such an oligomer. (\textit{Lower}) Amyloid-like fibrils formed by peptides in the $\beta$-sheet-competent form. For visual clarity, the red attractive patch in the $\beta$-form is depicted spanning the whole length of the spherocylinder body.}
\end{figure*}\label{Fig1}

\begin{figure*}[t!]
\center
\includegraphics[width=0.7\textwidth]{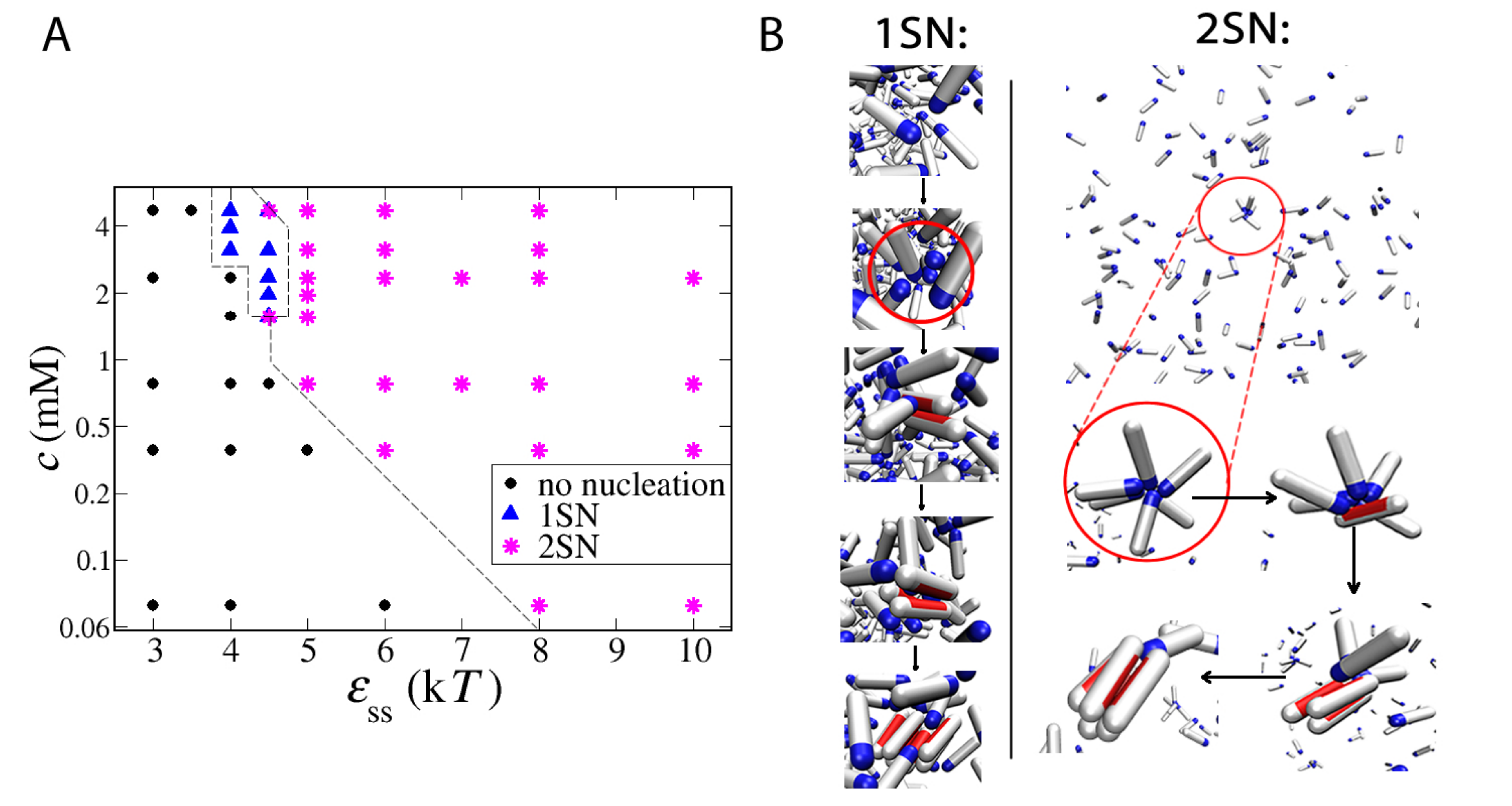}
\caption{Mechanism of amyloid nucleation. (\textit{A}) Tentative phase diagram. At low concentrations and low interpeptide interactions no nucleation was observed (black circles). The one-step mechanism is possible only at low interpeptide interactions and high concentrations (blue triangles). In any other case the nucleation proceeded through the two-step mechanism (magenta asterisks). All points are collected for $\epsilon_{s\beta}=\epsilon_{ss}+1\textrm{k}T$. The dashed lines are a guide to the eye. (\textit{B}) Pre- and postnucleation snapshots for two points in the phase diagram. (\textit{Left}) 1SN case, $c=1.8m$M, $\epsilon_{ss}=4.5\textrm{k}T$. The red circle zooms into the area where nucleation happens. (\textit{Right}) 2SN case at low peptide concentration, $c=72\mu$M, $\epsilon_{ss}=8\textrm{k}T$. The nucleating oligomer is circled in red in the first snapshot and magnified in the rest of the snapshots.}
\end{figure*}\label{Fig2}

\begin{figure*}[t!]
\centering

\center
\subfigure[]{\label{fig3a}\includegraphics[width=0.45\textwidth]{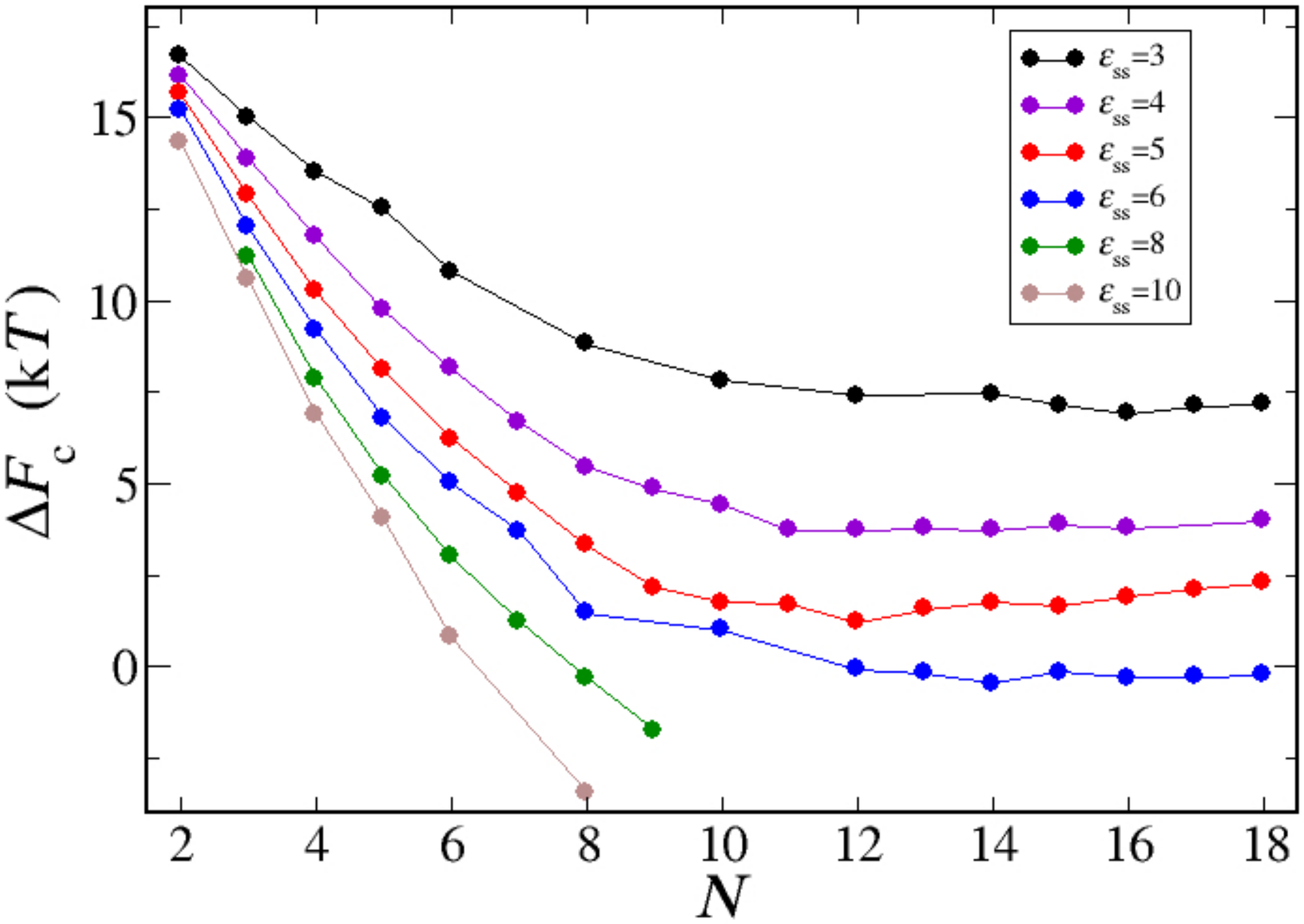}}
\subfigure[]{\label{fig3b}\includegraphics[width=0.45\textwidth]{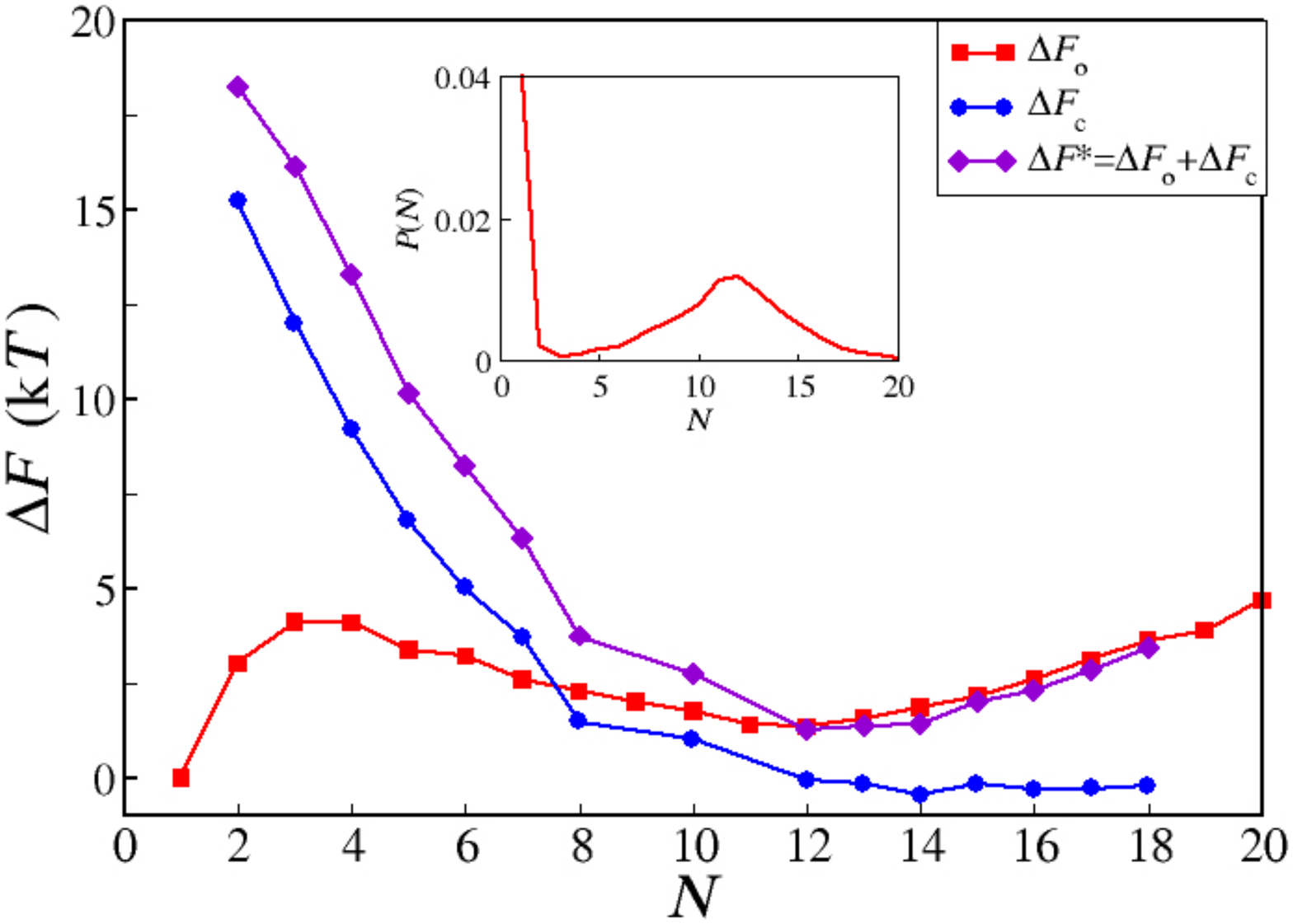}}
 \caption{(\textit{A}) The free energy barrier $\Delta F_c$ for conversion of one random-coil peptide into the $\beta$-form within a disordered oligomer of the size $N$, for various values of interpeptide interaction. From top to bottom $\epsilon_{ss}=3,4,5,6,8, 10$ $\textrm{k}T$. In all cases $\epsilon_{s\beta}=\epsilon_{ss}+1\textrm{k}T$ (\textit{B}) Partition of the free energy barrier for nucleation from solution. Red line and red squares: Free energy for oligomerization, $\Delta F_o(N)$, obtained from the oligomer-size distribution $P(N)$ (shown in \textit{inset}) at $c=2.17$mM and $\epsilon_{ss}=6\textrm{k}T$. The blue line and blue circles show the free energy for conversion of one peptide into the $\beta$-prone state within an oligomer of size $N$, $\Delta F_c(N)$, for  $\epsilon_{ss}=6\textrm{k}T$ and $\epsilon_{s\beta}=7\textrm{k}T$. Note that the blue line in \textit{B} corresponds to the blue line in \textit{A}. Violet line and violet diamonds: Free energy barrier for formation of a critical nucleus from solution, $\Delta F^*$, at $c=2.17$mM and $\epsilon_{ss}=6\textrm{k}T$, obtained as $\Delta F^*=\Delta F_o(N)+\Delta F_c(N)$. 
 }
\end{figure*}

\begin{figure*}[t!]
\centering
\center
\subfigure[]{\label{fi4a}\includegraphics[width=0.45\textwidth]{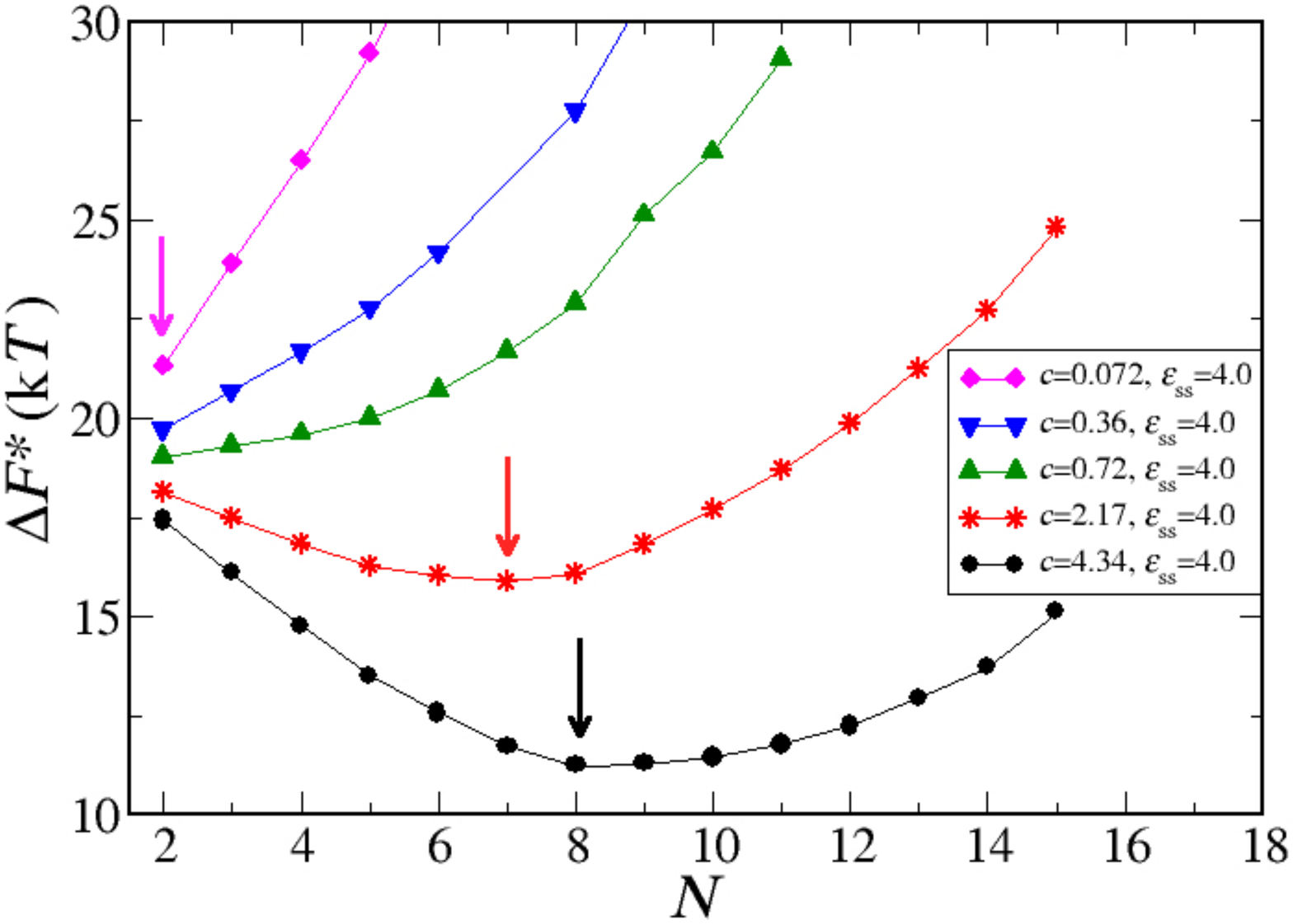}}
\subfigure[]{\label{fig4b}\includegraphics[width=0.45\textwidth]{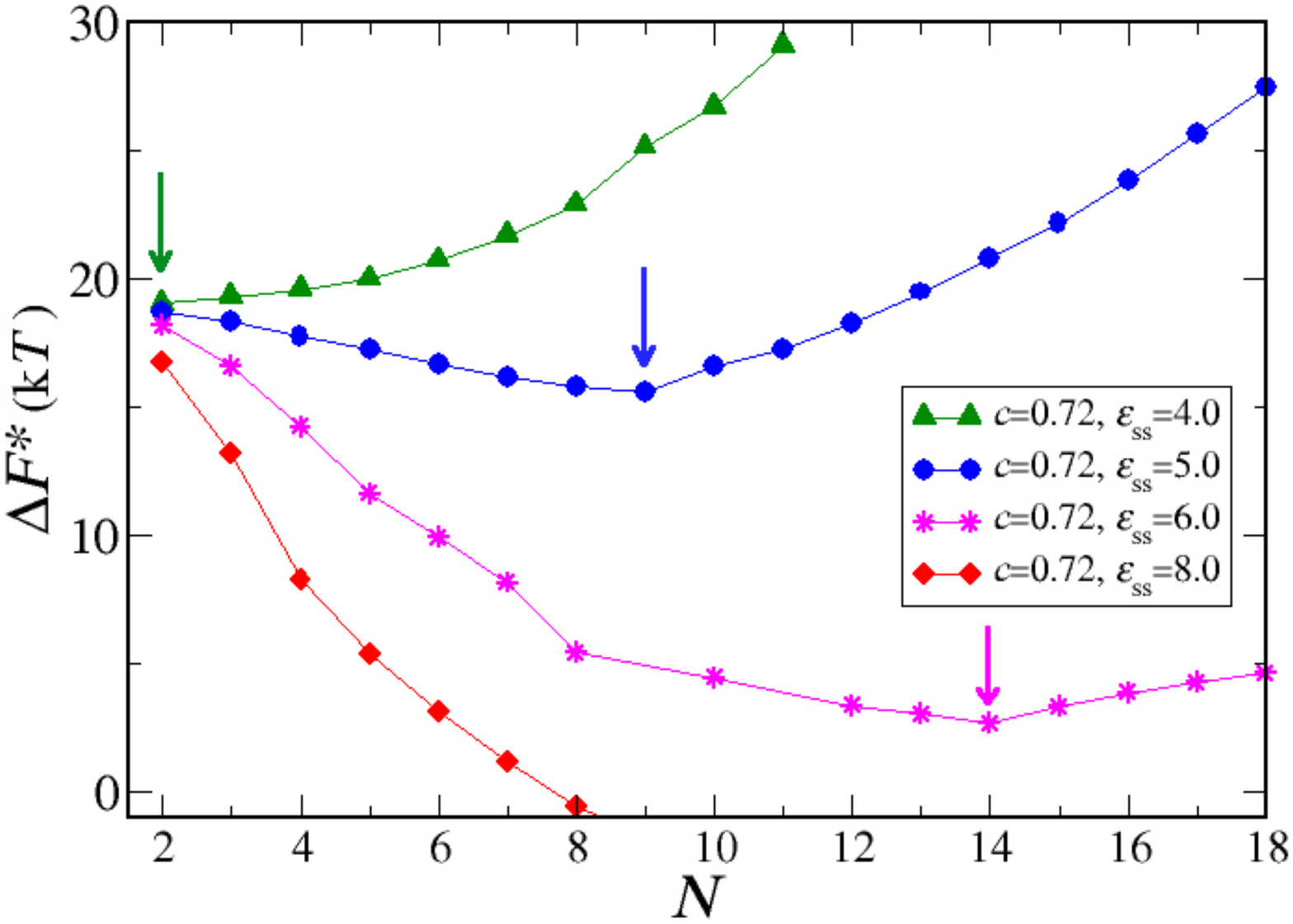}}
 \caption{ Free energy barrier for nucleation from solution. (\textit{A}) For constant interpeptide interaction and various peptide concentrations, $\epsilon_{ss}=4\textrm{k}T$ and all peptide concentrations are expressed in mM. (\textit{B}) For various interpeptide interactions (expressed in $\textrm{k}T$)and constant peptide concentration $c=0.72$mM. In all cases $\epsilon_{s\beta}=\epsilon_{ss}+1\textrm{k}T$. Arrows indicate critical nucleus sizes for several chosen parameters.}
\end{figure*}

\clearpage
\begin{center}

{\Large \textbf{Supporting Information}}

\end{center}
\bibliographystyle{unsrt}

\section*{I. Coarse-grained Simulations: Methods}

Each spherocylinder is $\sigma=2nm$ wide and $L=4\sigma=8nm$ long. The hard core repulsion forbids for any distance between any two spherocylinders to be smaller than $\sigma$. The interaction between  two  peptides in the soluble form is implemented as:
\begin{equation}
V_{ss}(r)=
  \begin{cases}
 -\epsilon_{ss}\left(\dfrac{\sigma}{r}\right)^6  & \text{if }  r \leq 1.5\sigma\\
      0     & \text{if } r > 1.5\sigma
  \end{cases}
\end{equation}\label{SI1}
where $r$ is the distance between the centers of the attractive tips located at the spherocylinders' ends. An attractive patch is added only at one spherocylinder pole to ensure formation of finite aggregates like those observed in experiments. This potential is reminiscent of a coarse-grained model for lipids~\cite{brown}. Because the center of the patch is only at one tip, there is an effective orientational dependence that is controlled by the cut-off of the tip-to-tip attraction. This potential drives the formation of spherical-like micelles, where tips of participating peptides are in contact in the micelle center. There is no other explicit angular dependence, and any angle between peptides in the $s$-state whose attractive tips are in contact are equally favourable.  

The attractive side-patch of the $\beta$-sheet forming configuration is $L_p=0.6L$ long and spans an angle of $180^\circ$. If two patches face each other their interaction is:
\begin{equation}
V_{\beta \beta}(r)=
  \begin{cases}
 -\epsilon_{\beta \beta} cos^2(\phi)-\epsilon_{\beta \beta} \left(\dfrac{\sigma}{r}\right) & \text{if } d\leq 1.5\sigma\\
      0     & \text{if } d > 1.5\sigma
  \end{cases}
\end{equation}\label{SI2}

where $\phi$ is the angle between the axes of the particles, $d$ is the shortest distance between the axes of the patches, and $r$ is distance between the centers of the patches. The first term controls that peptides in the $\beta$-forms pack parallel to each other, mimicking the hydrogen-bond interactions between $\beta$-sheets, while the second term ensures compactness of the fibrils~\cite{fitzpatrick, serpell,eisenberg}. In this study we choose $\epsilon_{\beta\beta}=30\textrm{k}T$~\cite{berkowitz,welland}. Aggregation of patchy-spherocylinders has been studied in details in our previous work~\cite{vacha1}.
The cross-interaction between the soluble and the $\beta$-sheet-forming configuration is designed like:
\begin{equation}
V_{s\beta}(r)=
  \begin{cases}
 -\epsilon_{s\beta} & \text{if } d < 1.5\sigma\\
      0     & \text{if } d > 1.5\sigma
  \end{cases}
\end{equation}\label{SI3}
where $d$ is the shortest distance between the centre of the attractive tip and the axis of the $\beta$-patch. This interaction is also effectively orientation-dependent, which is controlled by the cut-off of the potential, and the angular-width of the $\beta$-patch.\par
We simulate the system of 600 peptides in a cubic box at volume fractions 0.06, 0.05, 0.04, 0.03, 0.025, 0.02, 0.01, 0.005 and 0.001, corresponding to peptide concentrations of 4.34 mM, 3.61 mM, 2.89 mM, 2.17 mM, 1.81 mM, 1.45 mM, 0.72 mM, 0.36 mM and 72 $\mu$M respectively. Each Monte Carlo run was at least $10^6$ steps long. \par
To obtain the free energy barrier $\Delta F_{c}(N)$ for conversion of one soluble peptide into the $\beta$-form within an oligomer of a size $N$, we restrain the centers of attractive tips of $N$ monomers to stay within a spherical space of radius $5\sigma$. In this way, monomers are kept in a micellar-like conformation, while having enough freedom to find their preferential orientation with respect to each other. A biasing harmonic potential is then applied, which ensures that the number of monomers in the $\beta$-form oscillates between 0 and 1. Regarding the value of the harmonic spring that is applied, we scanned a wide range of values in each case, until we found the one that equally well samples both sides of the barrier (the side where all peptides are in the $s$-state, and the one where exactly one peptide has converted into the $\beta$-state).
\par
The free energy change for oligomerisation is obtained from simulating 600 monomers in a soluble form and collecting the oligomer-size distribution $P(N)$. The free energy for formation of an oligomer of the size $N$ is then $\Delta F_{o}(N)=-logP(N)+F^0$, where the zero-energy level $F^0$ is attributed to free monomers in the solution.\par
Finally, the free energy barrier to nucleate from the solution is $\Delta F^*(N)=\Delta F_{o}(N)+\Delta F_{c}(N)$.\par

\vspace{4 mm}

\section*{II. Umbrella Sampling of A$\beta$1-42 Interactions: Methods}

The starting conformation of the atomistic simulation is based on the 3D model of the A$\beta$ protofibril consisting of five peptides (PDB 2BEG)\cite{Luehrs2005}. The residues 1 to 16 were added to the A$\beta$17-42 dimer from this structure using the building tool in Pymol in a disordered conformation with randomly chosen $[\varphi,\psi]$ dihedral values.\par
Simulations were performed using the Amber99SB-ILDN force field\cite{Piana2011} and the TIP3P water model\cite{Jorgensen1983} in the GROMACS 4.6.3 package\cite{Hess2008} with Ewald summation for long-range electrostatic interactions\cite{Essmann1995}. The LINCS algorithm\cite{Hess1997} was used to constrain all bond lengths to their equilibrium values and a time step of 2.0 fs was set. \par
To describe the $\beta$-$\beta$ interaction, dihedral restraints were applied on the sequence 17-42 forming the hairpin of the dimer, with a force constant of 200 kJmol$^{-1}$rad$^{-2}$. These restraints were used to preserve the fibril-competent state, and prevent it from unfolding as the monomers are separated and the inter-chain hydrogen bonds are broken. In order to generate the disordered starting conformation for the other two types of interactions (soluble-$\beta$ and soluble-soluble), a molecular dynamics simulation was performed at 500~K from the built conformation of A$\beta$1-42 for 1000 ps. Each A$\beta$ dimers ($\beta$-$\beta$, s-$\beta$, s-s) were solvated in a periodic box of 1,512~nm$^3$ with approximately 50,300 water molecules and a salt concentration of 0.1 M.
All three systems were subject to a minimization using the steepest descent algorithm and Molecular Dynamics (MD) simulations at constant volume and temperature for 100 ps, with position restraints on all heavy atoms, in order to equilibrate the solvent molecules. This step was followed by 100 ps of MD simulations in the Canonical ensemble without positional restraints. The systems were then simulated for a last step of 100-ps long MD in the isothermal-isobaric ensemble. The Nos\'{e}-Hoover thermostat\cite{Nose1984,Hoover1985} was applied for temperature control and the Parrinello-Rahman barostat\cite{Parrinello1981} for pressure control. 
The temperature and pressure of the simulations were set to 300 K and 1 bar, respectively. \\
The resulting unsolvated conformation of A$\beta$ peptides were then used to prepare the umbrella simulations starting conformations. The chains were separated along the z-axis of the center of mass vector between them. The initial distances were of 0.85, 1.2 and 2 nm for the $\beta$-$\beta$, soluble-$\beta$ and soluble-soluble conformations, respectively. All starting states were then solvated in the same conditions described above in a periodic of box of 1,512~nm$^3$, followed by a 100-ps position restrained simulation and a total of 200 ps of unrestrained equilibration. \\
Umbrella sampling simulations were then performed for 10ns using a spring constant of 3000 kJmol$^{-1}$nm$^{-2}$ for each three sets of A$\beta$ conformations, using as a reaction coordinate the center of mass separation along the z-axis. It is important to note that the shortest separation for the umbrella simulations of the disordered conformation are limited by the radius of gyration for each chain, preventing the shortest sampled distance to be smaller than the sum of the respective radii of gyration. 
The Potentials of mean force (PMFs) were calculated using the Weighted Histogram Analysis Method (WHAM) as implemented in GROMACS\cite{Hub2010} without the first ns for equilibration purposes.\\
The number of atomic hydrophobic and polar contacts and hydrogen bonds were computed using GROMACS analysis tools.
The same simulation protocol was used to perform the umbrella simulations of the A2V variant
of A$\beta$42.

\clearpage
\newpage

\begin{figure*}[t!]
\centering
\includegraphics[width=0.95\textwidth]{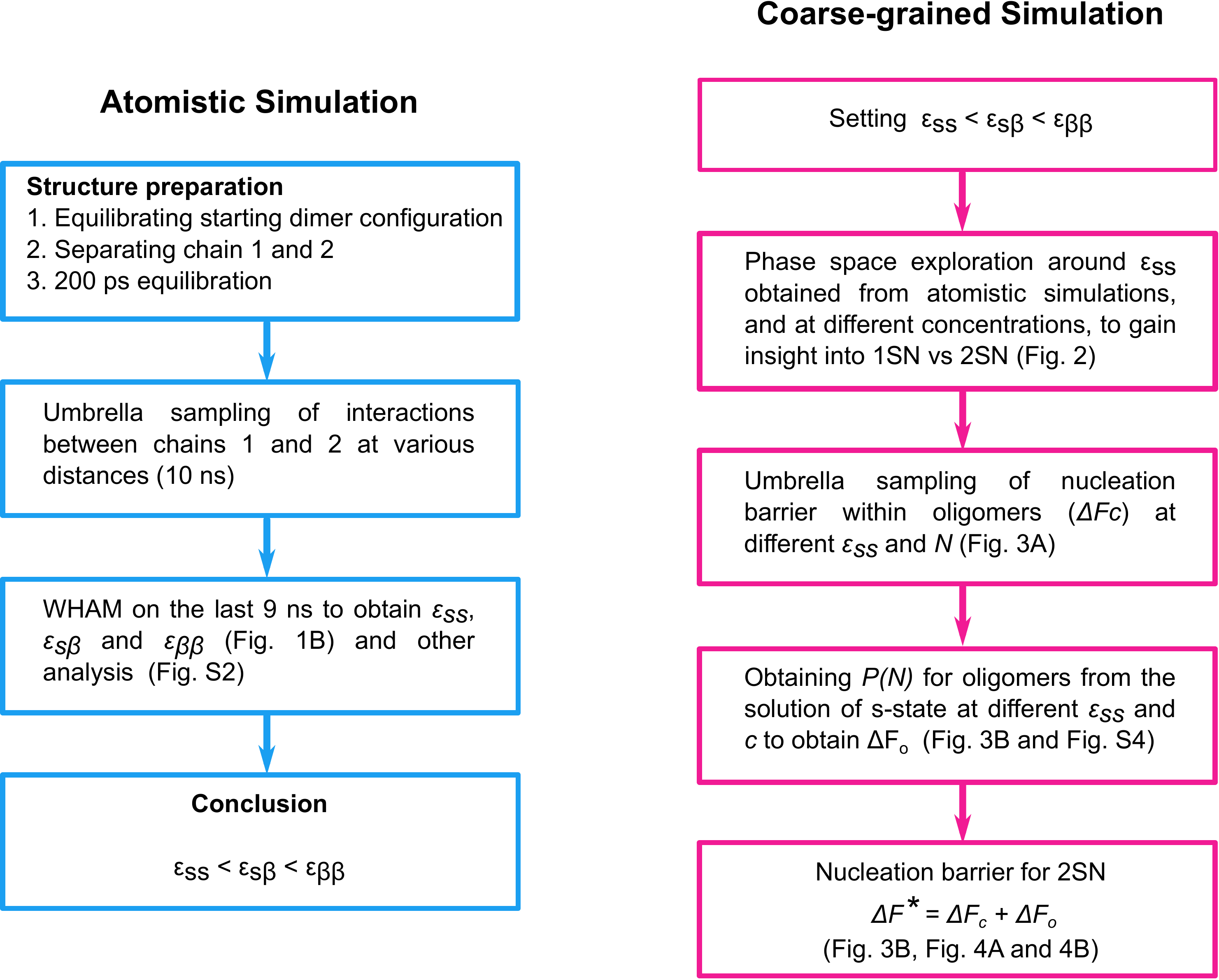}
\caption{Chart of the computational workflow carried out in this paper.}
\end{figure*}

\begin{figure*}[t!]
\centering
\center
\includegraphics[width=0.7\textwidth]{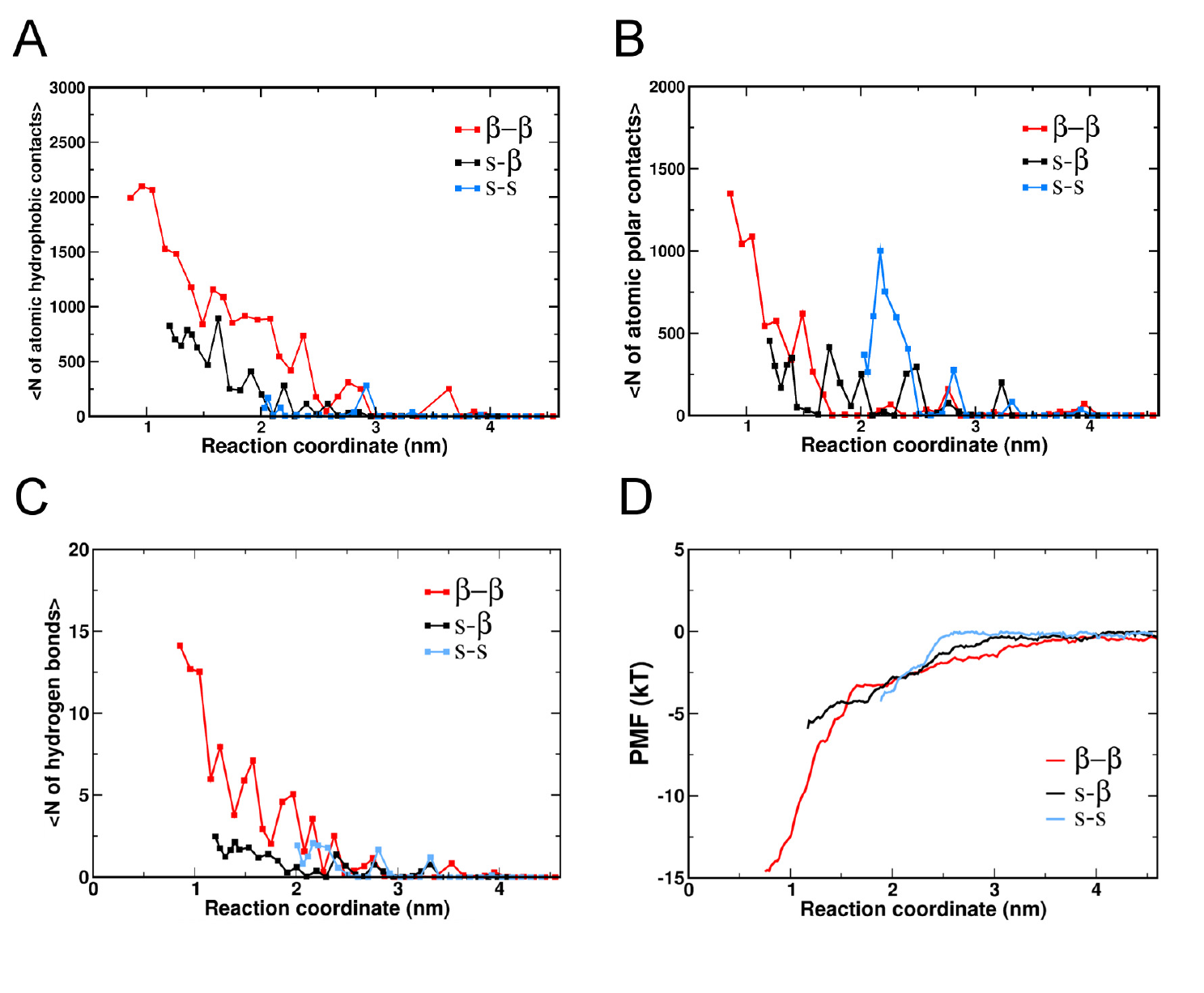}
\caption{The average number of contacts between (A) atoms of hydrophobic residues, (B)
atoms of polar residues and (C) the average number of hydrogen bonds, 
calculated for each umbrella simulation as a function of the reaction coordinate for all three 
pair-interactions: $\beta-\beta$ (red line and red squares), 
soluble-$\beta$ (black line and black squares), soluble-soluble (blue line and blue squares). 
(D) Potential of mean force (PMF) for the pair interactions in the A2V mutant of the A$\beta$1-42 system, 
illustrating the interaction between two peptides in the $\beta$-form (red line), 
the interaction between the soluble peptide and the peptide kept 
in the $\beta$-prone form (black line) and the interaction between the soluble peptides (blue line).}
\end{figure*}

\begin{figure*}[t!]
\centering
\includegraphics[width=0.45\textwidth]{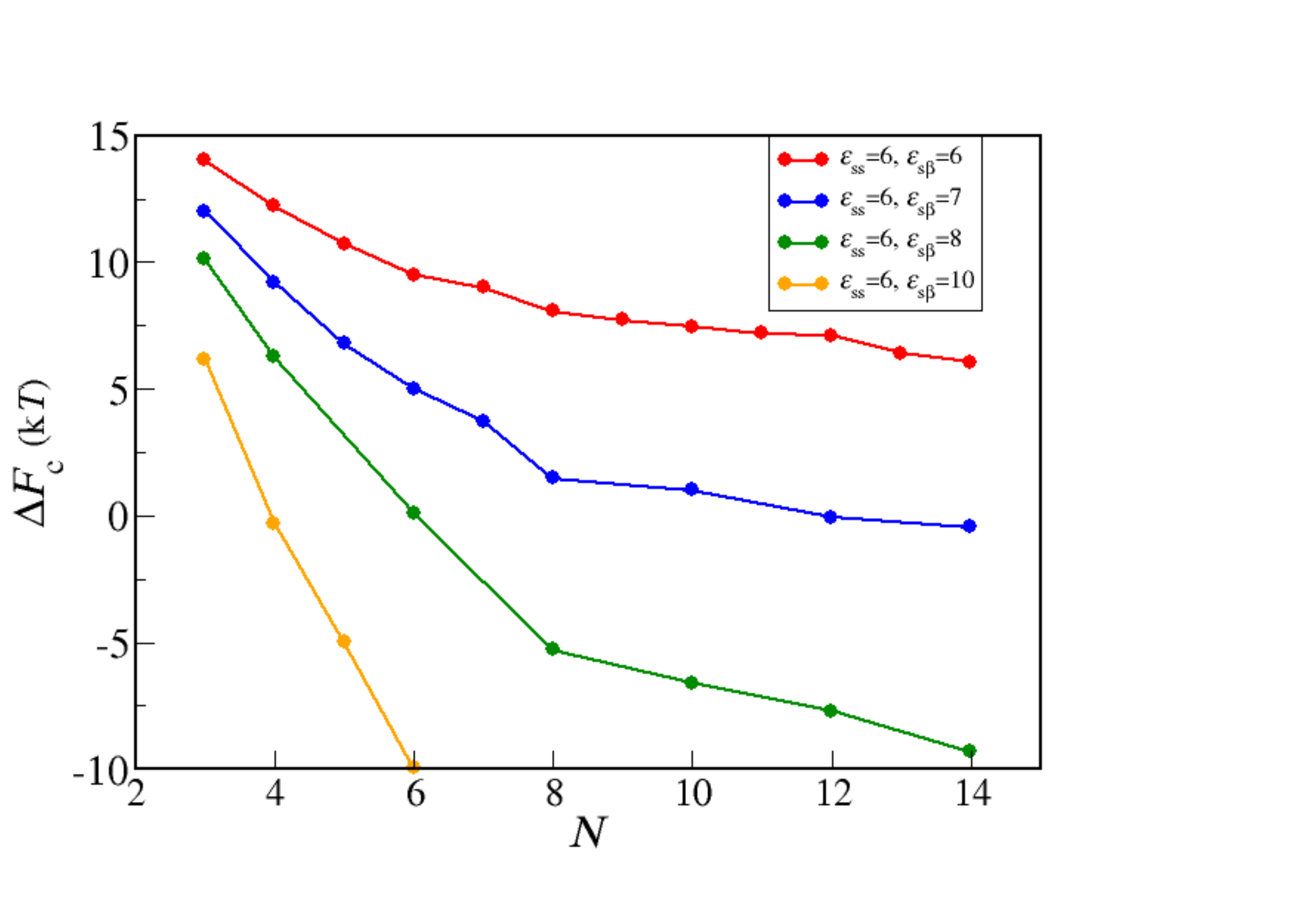}
\caption{The free energy barrier $\Delta F_c$ for conversion of one random-coil peptide into the $\beta$-form within a disordered oligomer of the size $N$, for various values of $\epsilon_{s\beta}$ interaction parameter. From top to bottom $\epsilon_{s\beta}=6,7,8, 10\textrm{k}T$. The soluble-soluble interaction is kept constant at $\epsilon_{ss}=6\textrm{k}T$.}
\end{figure*}

\begin{figure*}[t!]
\centering
\includegraphics[width=0.45\textwidth]{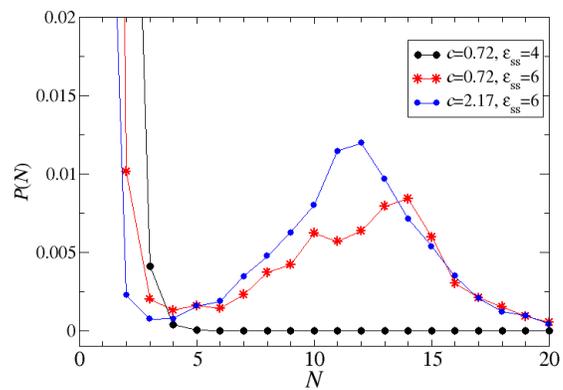}
\caption{Oligomer size distribution $P(N)$ for three combinations of $c$ and $\epsilon_{ss}$. Note that for all three cases most of the peptides are in the monomeric form, and the system is below the critical micelle concentration.}
\end{figure*}

\end{document}